\documentclass[twocolumn,floats,prl,aps,showpacs]{revtex4-1}
\usepackage{float}
\usepackage{setspace}
\usepackage{xkeyval}
\usepackage{rotating}
\usepackage{amsmath}
\usepackage{color}
\usepackage{graphicx}
\usepackage{courier}
\usepackage[sort&compress]{natbib}

\newcommand{\qq}{{\bf q}}

\newcommand{\kk}{{\bf k}}

\newcommand{\be}{\begin{equation}}
\newcommand{\ee}{\end{equation}}
\newcommand{\ben}{\begin{equation*}}
\newcommand{\een}{\end{equation*}}
\newcommand{\bea}{\begin{eqnarray}}
\newcommand{\eea}{\end{eqnarray}}
\newcommand{\bean}{\begin{eqnarray*}}
\newcommand{\eean}{\end{eqnarray*}}

\def\efield{\boldsymbol{\cal E}}

\newcommand{\cnrs}{Univ. Grenoble Alpes/CNRS, Institut N\'eel, F-38042 Grenoble, France} 
\newcommand{\qub}{School of Mathematics and Physics, Queen's University Belfast, Belfast BT7 1NN, Northern Ireland, UK}
\newcommand{\coimbra}{Centre for Computational Physics and Physics Department, University of Coimbra, Portugal}

\begin{document}
\title[SHG in h-BN and MoS$_2$: role of e-h interaction]{Second Harmonic Generation in h-BN and MoS$_2$ monolayers:\\ the role of electron-hole interaction}
\author{M. Gr\" uning$^{1,2}$ and C. Attaccalite$^3$}
\affiliation{$^1$\qub \\ $^2$\coimbra \\ $^3$\cnrs}

\begin{abstract}
In this letter we show by means of first principle numerical simulations that  electron-hole interaction significantly contributes to the second-harmonic generation spectrum  of h-BN or MoS$_2$ monolayers.
Specifically, it doubles the signal intensity at the excitonic resonances with respect to the contribution from independent electronic transitions. This result hints that the intensity of second-harmonic signal of those materials can be tuned by changing the dielectric screening, that controls the strength of the electron-hole interaction. 
\end{abstract}           
\pacs{78.20.Bh,78.67.-n,42.65.Ky}

\maketitle

\emph{Introduction.}~
Optical properties of two-dimensional (2D) semiconducting crystals, and in the specific of hexagonal boron nitride (h-BN) and  MoS$_2$  monolayers, have been object of an intense research in the past years (e.g. Refs~\cite{doi:10.1021/nl903868w,PhysRevLett.105.136805,watanabe,Watanabe2004}). Several studies have investigated the absorption and photoluminescence spectra and possible applications to optoelectronics (for a review see Ref.~\cite{doi:10.1021/nn403159y}).
  
Significative advances in the knowledge of linear optical properties of h-BN and MoS$_2$ have been possible also thanks to ab-initio studies that contributed both through the interpretation of experimental results (e.g. Refs.~\cite{molina2013effect,PhysRevLett.100.189701}) and the envisagement of possible applications (e.g. Ref.~\cite{doi:10.1021/nl401544y}). Key in those studies has been the inclusion of the electron-hole interaction, essential to capture the excitonic features---particularly strong due to the geometric confinement and the weak dielectric screening---that characterize the optical response of 2D crystals~\cite{scholes2006excitons}. 

More recently, there has been a surge of interest also for nonlinear optical properties of these materials. Several experimental studies~\cite{kumar2013second,doi:10.1021/nl401561r,PhysRevB.87.201401} show that h-BN and MoS$_2$ monolayers have a remarkable second-harmonic generation (SHG) hinting potential applications to nonlinear optical devices. Being sensitive to the stacking, orientation and number of layers, SHG has been proposed and already used as noninvasive optical probe to characterize h-BN and MoS$_2$ films~\cite{kumar2013second,doi:10.1021/nl401561r}. On the other hand, as for linear optical properties, it is important to support or even guide experiments through accurate and reliable numerical simulations. As an example, for the MoS$_2$ monolayer depending on the study the experimental estimate for the SHG varies by 4 orders of magnitude~\cite{kumar2013second,doi:10.1021/nl401561r,PhysRevB.87.201401}.

Unfortunately, in contrast with the case of linear optical properties calculations of the non-linear optical response still remain a challenge.
With few isolated exceptions~\cite{Leitsmann2005,Chang2002,PhysRevB.82.235201}, large part of calculations of nonlinear optical properties, and in the specific of SHG in 2D-crystals~\cite{guo2005second,margulis2013optical,2013arXiv1310.0674T}, employ the independent particle approximation (IPA) which is inadequate for low dimensional systems, where it is expected that the strongly bound excitons significantly modify the SHG. 

One of the main obstacles for the inclusion of electron-hole interaction in calculations of nonlinear optical properties actually comes from  implementing the expression for the correlated nonlinear susceptibility in terms of the electronic structure of the systems, with the complexity of the expression growing with the nonlinear order. For example, within Many-body perturbation theory the diagrams that enter in the calculation of the second and third order susceptibilities,  are so intricate that their implementation becomes awkward if not impracticable (see for instance Fig.~3 of Ref.~\cite{PhysRevB.80.165318}). Even within the ``simpler'' time-dependent density-functional theory, equations for the second order optical response has been solved only for particular approximations for the correlation functional~\cite{PhysRevB.82.235201}, or have been limited to the static response~\cite{kirtman:1294}.  

In a recent work~\cite{nloptics2013}, we propose to avoid the direct calculation of the nonlinear optical susceptibilities, and use instead a real-time approach.  In such an approach the optical susceptibilities are obtained from the time propagation of ``simpler'' objects such as the single particle Green's function, the density matrix or the density. Correlation effects then are included easily as an operator into the time-dependent (effective) Hamiltonian.
This approach has been already successfully implemented and used for example within time-dependent density-functional theory, but usually limited to finite systems, such as atoms, molecules or clusters~\cite{takimoto:154114}. Our approach is instead designed to treat periodic systems, such as crystals, and is based on an approximation for the electron-hole interaction~\cite{strinati} derived from Many-body perturbation theory that proved to be successful for linear optical properties (e.g. Refs.~\cite{molina2013effect,PhysRevLett.100.189701}). 

Here we apply this approach to calculate and analyze the contribution of electron-hole interaction on the SHG spectra of h-BN and MoS$_2$ monolayers. For both materials we disclose the signature of bound excitons and show that excitonic effects not only significantly modify the shape of the spectrum with respect to the IPA, but strongly enhance its intensity.
In the conclusions we comment how this finding may open the possibility of engineering the SHG signal in these materials.

\emph{Computational Methods.}~
The main equation in our real-time approach to nonlinear optical properties, is the equation of motion for the time-dependent valence states $| v_{\mathbf {k},m} \rangle$ 
\bea
i\hbar  \frac{d}{dt}| v_{\mathbf {k},m} \rangle &=& (H^0_{\mathbf k} + \Delta H_{\mathbf k}) | v_{\mathbf {k},m} \rangle + V_h[\Delta \rho]\notag \\ &+& \Sigma_{\text{SHF}} [\Delta \gamma] + \efield | \partial_\kk v_{\mathbf {k},m} \rangle \label{tdbse_shf}.
\eea 
where $\Delta \rho(\mathbf r) = \rho(\mathbf r;t)-\rho(\mathbf r;t=0)$ and $\Delta \gamma = \gamma(\mathbf r,\mathbf r';t) - \gamma(\mathbf r,\mathbf r';t=0)$ are respectively the variation of the electronic density and of density matrix induced by the external field $\efield$. The last term on the r.h.s. of Eq.~\eqref{tdbse_shf} describes the coupling with the external field. We treat this term within Modern Theory of Polarization~\cite{RevModPhys.66.899} in the extension to dynamical polarization proposed by Souza et al.~\cite{souza_prb} that we recently implemented in an \emph{ab-inito} framework~\cite{nloptics2013}.

The rest of the terms on the r.h.s. corresponds to different approximations for the effective Hamiltonian. 
The first term, $H^0_\kk$ is the unperturbed mean-field Hamiltonian, for which we choose the Kohn-Sham one~\cite{PhysRev.140.A1133}, and corresponds to the IPA. The second term, $\Delta H_\kk$ is the so-called scissor operator, that corrects the band structure of $H^0_\kk$, to provide the quasiparticle band-structure. Here the scissor is evaluated within the $GW$ approximation~\cite{aulbur1999quasiparticle} and the corresponding approximation referred as IPA+$GW$ corrections. This term is responsible for the local-field effects~\cite{PhysRev.126.413} originating from system inhomogeneities. The third term is the Hartree~\cite{attaccalite} potential. By truncating  Eq.~\eqref{tdbse_shf} at this level one obtains the time-dependent Hartree (TDH) approximation. 
The next term in Eq.~\eqref{tdbse_shf} is the screened Hartree-Fock (SHF) self-energy $\Sigma_{\text{SHF}}$, that accounts for the electron-hole interaction~\cite{strinati}, and is written as $\Sigma_{\text{SHF}}[\Delta \gamma] = W(\mathbf r,\mathbf r') \Delta\gamma(\mathbf r,\mathbf r';t)$ where the static screened Coulomb interaction $W(\mathbf r,\mathbf r')$ is calculated in random phase approximation keeping the screening fixed to its zero-field value (for details see Ref.~\cite{attaccalite}). Approximation at this level is referred as time-dependent SHF (TDSHF). Note that, within Green's function theory,  the linear response limit of the full Eq.~\eqref{tdbse_shf} is equivalent to the solution of the Bethe-Salpeter equation~\cite{strinati} in static ladder approximation on top of the $G_0W_0$ quasi-particles band structure~\cite{aulbur1999quasiparticle}.
   
In order to calculate the non-linear optical response to an external field, we choose a monochromatic (sinusoidal) electric field in Eq.~\eqref{tdbse_shf}, and calculate the time-dependent polarization as:
 \begin{equation}
 \mathbf P_\parallel = -\frac{ef}{2 \pi v} \frac{\mathbf a}{N_{\kk_\perp}} \sum_{\kk_\perp} \mbox{Im log} \prod_{\kk_\parallel}^{N_{\kk_\parallel}-1}\ \mbox{det} S(\kk , \kk + \mathbf q_\parallel), \label{berryP} 
 \end{equation}
 where $\mathbf P_\parallel$ is the polarization along the lattice vector $\mathbf a$, $v$ the unit cell volume, $S(\kk , \kk + \mathbf q_\parallel) $ is the overlap matrix between the time-dependent valence states $|v_{\kk,n}\rangle$ and $|v_{\kk + \qq_\parallel,m}\rangle$, $N_{\kk_\parallel}$ and $N_{\kk_\perp}$ are respectively the number of $k$ points along and perpendicular to the polarization direction, and $\mathbf q_\parallel = 2\pi/(N_{\kk_\parallel} {\mathbf a})$.
Then, the second harmonic coefficient is extracted from the power series of total polarization $P=\chi^{(1)} \efield + \chi^{(2)} \efield \efield + ....$  as explained in more details in Ref.~\cite{nloptics2013}.

We apply the method here reviewed to of h-BN~\cite{hbnparameters} and MoS$_2$~\cite{mos2parameters} monolayers. 
Valence states are expanded in a plane-wave basis set and the isolated monolayers are simulated by a slab supercell approach with large inter-sheet distance. Numerical details can be found in Refs.~\cite{intequation,hbnparameters,mos2parameters}.

\begin{figure}[ht]
\centering
\vspace{-0.2cm}
\includegraphics[width=0.48\textwidth]{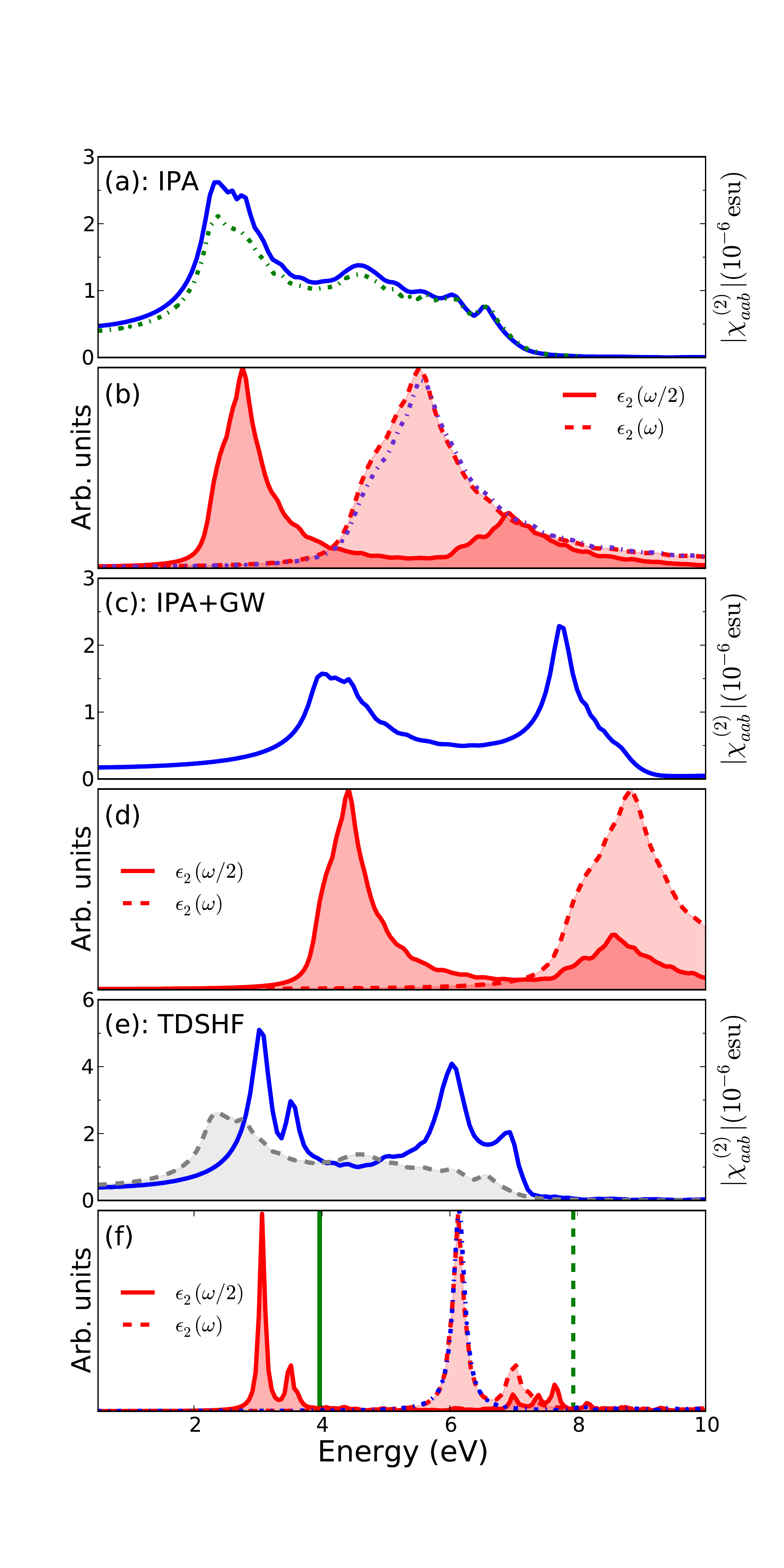}
\vspace{-0.5cm}
\caption{\footnotesize{SHG spectra for the h-BN monolayer at different levels of theory [Eq.~\eqref{tdbse_shf}]: (a) IPA (blue continuous line) and TDH (green dashed line); (c) IPA + GW correction (blue continuous line); (e) TDSHF (blue continuous line) and IPA (grey dashed line). The imaginary part of the dielectric constant at both $\omega/2$ (red continuous line) and $\omega$ (red dashed line) is reported in (b), (d) and (f) for IPA, TDH and TDSHF respectively. 
The vertical lines represent the $GW$ fundamental gap (green dashed line) and half of the $GW$ fundamental gap (green continuous line). \label{absX2bn}}}
\vspace{-0.2cm}
\end{figure} 

\emph{h-BN monolayer.}~
h-BN is a transparent insulating material with a large band gap of about $6~eV$. Its absorption spectrum is dominated by strong bound excitons, nearly independent from the layers arrangement~\cite{PhysRevLett.96.126104,PhysRevLett.100.189701}. The h-BN monolayer inherits all these properties from its bulk counterpart.

In Fig.~\ref{absX2bn} we report the calculated absolute value of $\chi^{(2)}_{aab} (\omega)$, the only independent in-plane component of $\chi^{(2)} (\omega)$ ($a$ and $b$ are the in-plane cartesian directions) at different levels of approximation. Assignment of the peak is done by comparison  with the imaginary part of the independent particle dielectric constant $\epsilon_2$ [Fig.~\ref{absX2bn}(b,d,f)].
At IPA level [Fig.~\ref{absX2bn}(a)], the SHG presents a peak at $2.3~eV$ and a broad structure between $4 - 7 eV$, corresponding respectively to two-photon  and one-photon resonances with $\pi \to \pi^*$ transitions, with contributions around $7 eV$ of two-photon resonances with $\sigma \to \sigma^*$ transitions. The IPA level of theory is the one usually employed in theoretical calculations of SHG. Results for the h-BN monolayer were previously obtained by Guo and Lin~\cite{guo2005second} and are in good agreement with our calculations (see also Table~\ref{tab1}). In the following we show how effects beyond the IPA---that is the additional terms in Eq.~\eqref{tdbse_shf}---modify the SHG spectrum.
 
We start by adding crystal local field effects, included at the TDH level [Fig.~\ref{absX2bn}(a)]. Because of the weak in-plane inhomogeneity of the h-BN, local field effects are small and reduce by about 20\% the peak at $2.3~eV$. 
Note however they are larger than in the absorption spectrum for which the effect is negligible [Fig.~\ref{absX2bn}(b)]. 
Next we consider the renormalization of the band structure by quasiparticle corrections within the $GW$ approximation (IPA+$GW$) [Fig.~\ref{absX2bn}(c)]. For h-BN this renormalization can be safely approximated by a rigid shift of the conduction bands~\cite{hbnparameters}. Differently from the absorption spectrum [Fig.~\ref{absX2bn}(d)], the SHG is not simply shifted by $GW$ corrections, but its shape changes remarkably as a consequence of the more involved poles structure of the second order susceptibility~\cite{PhysRevB.82.235201,hughes1996calculation}.
In fact, the IPA+$GW$ shows two peaks: the first at about $4~eV$ is the shifted two-photon $\pi \to \pi^*$  resonances peak which is attenuated by 40\% with respect to IPA  [Fig.~\ref{absX2bn} (a)]; the second very pronounced peak at about $8~eV$ comes from the interference of  $\pi \to \pi^*$  one-photon resonances and  $\sigma \to \sigma^*$ two-photon resonances.  

Finally, in Fig.~\ref{absX2bn}(e) we consider the full Hamiltonian in Eq.~\eqref{tdbse_shf}. In particular we add the SHF term that introduces an attractive interaction between the excited electrons and holes~\cite{strinati}. The SHG spectrum presents four sharp and strong peaks and its onset is red-shifted by about $1~eV$ with respect to the the IPA+$GW$ [Fig.~\ref{absX2bn} (c)]. The two couples of peaks can be identified respectively as the two- and one-photon resonances with the excitons at $6$ and $7~eV$.
Figure~\ref{absX2bn}(c) also emphasizes the striking difference with respect to IPA, and shows that TDSHF is twice as strong as IPA at the exciton resonances. 
In Table~\ref{tab1} we report  the value of the second optical susceptibility at $\omega=0$, $\chi^{(2)}(\omega \to 0)$, extrapolated from the SHG  behavior at small frequencies.
Again, at the IPA level our result agrees with the one of Guo and Lin~\cite{guo2005second} within the error bar. Adding the effects beyond IPA, modifies the $\chi^{(2)}(\omega \to 0)$ value, and in particular within TDSHF we found a value smaller by about 10\% than at the IPA level.
Experimentally, Ref.~\cite{doi:10.1021/nl401561r} provides an estimate for the SHG at 1.53 eV (810 nm) (assuming an effective layer thickness of 3.3 \AA) of about $5\cdot 10^{-8}$ esu, one order of magnitude smaller than what we find, though a direct comparison is not possible since experiments measure second-harmonic signal of the monolayer on a substrate, while we evaluate the SHG of the monolayer in vacuum. 

\begin{center}
\begin{table}[h]
\vspace{-0.3cm}
\small
\begin{tabular}{c|cc|c|c|c}
\hline
$|\chi^{(2)}_{aab} (0)| $ (pm/V) & \multicolumn{2}{c|}{IPA} & TDH & IPA+G$_0$W$_0$ & TDSHF  \\
\hline
 h-BN &  41.2(7)  & [40.7] & 34.7(9) &  16.8(1)  & 36.8(3)  \\
\hline
\end{tabular}
\caption{$\omega\rightarrow 0$ limit of $\chi^{(2)}_{aab} (-2\omega,\omega,\omega)$ of the h-BN monolayer at different levels of the theory [Eq.~\eqref{tdbse_shf}]. As a comparison, for the IPA we report in square brackets also the value obtained in Ref.~\cite{guo2005second}.\label{tab1}}  
\end{table}
\vspace{-0.7cm}
\end{center}

\emph{MoS$_2$ monolayer.}~
MoS$_2$ differs from h-BN in several aspects. First, in MoS$_2$ an indirect-to-direct band gap transition occurs passing from the bulk to the monolayer due to the vanishing interlayer interaction.  Second, spin-orbit coupling plays an important role in this material, splitting the top valence bands, as visible from the absorption spectrum, presenting a double peak at the onset~\cite{PhysRevLett.105.136805}. Third, Mo and S atoms in the MoS$_2$ monolayer are on different planes resulting in a larger inhomogeneity than for the h-BN.

\begin{figure}[h]
\vspace{-0.5cm}
\includegraphics[width=.5\textwidth]{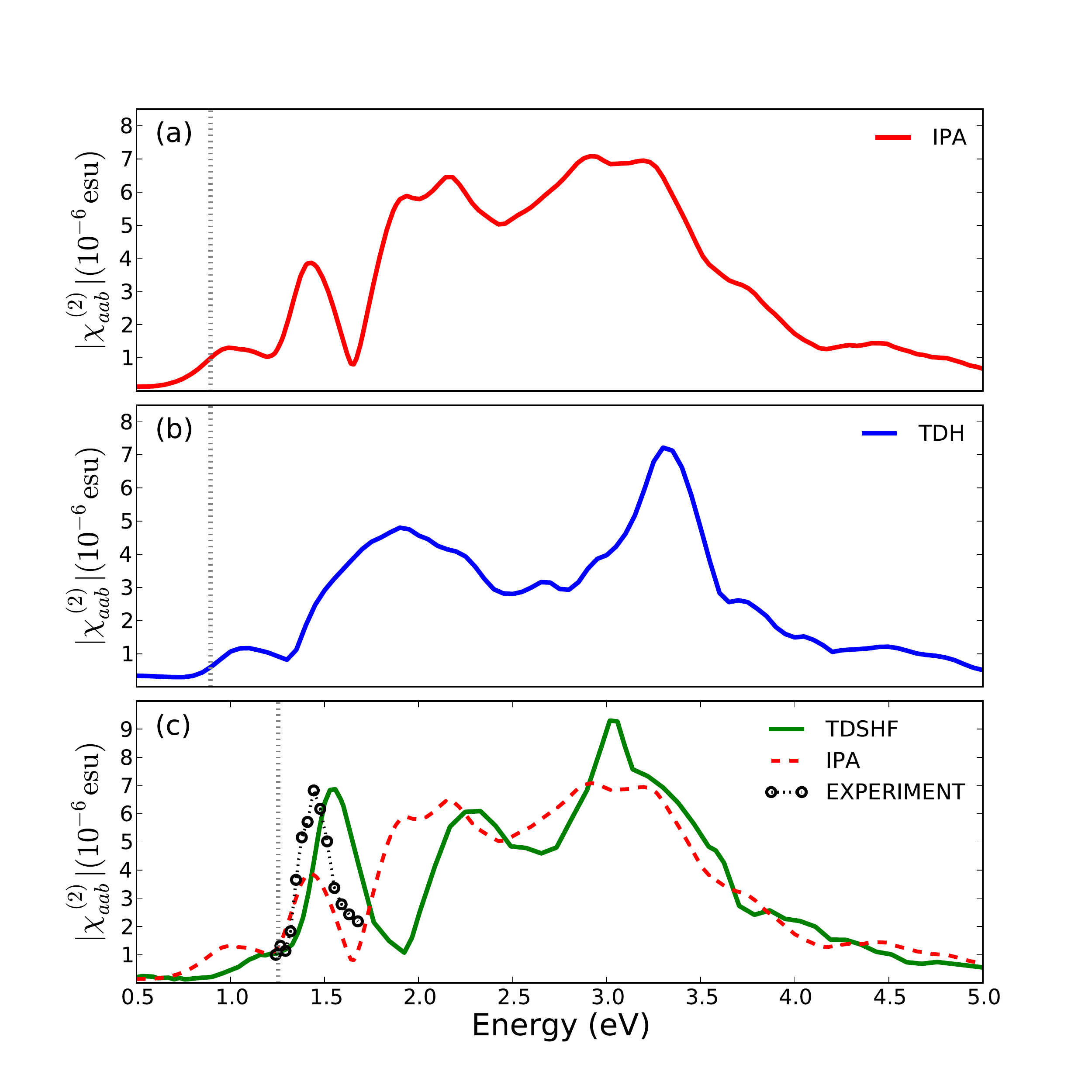}
\vspace{-0.7cm}
\caption{\footnotesize{SHG in the MoS$_2$ monolayer for different approximations: (a) IPA, (b) TDH and (c) TDSHF (green line). The latter is compared with IPA (red dashed line) and experimental results of Malard et al.~\cite{PhysRevB.87.201401} (black circles). The intensity of the experimental spectrum has been renormalized to match the intensity of the $~ 1.5 eV$ peak (see text). The dotted vertical lines show the energy of half of the Kohn-Sham band gap in (a) and (b), and of half of the $GW$ band gap in (c).}\label{fg:SHMoS2}}
\vspace{-0.3cm}
\end{figure}

Figure~\ref{fg:SHMoS2} presents the SHG spectra $|\chi^{(2)}_{aab}|$ of the MoS$_2$ monolayer at the different level of approximations of Eq.~\eqref{tdbse_shf}. At the IPA level [Fig~\ref{fg:SHMoS2} (a)], the SHG presents three main features: a small peak at 1~eV, which originates from two-photon resonances with transitions close to the minimum gap at the $K$ point; a larger peak around $1.5$~eV, which originates from two-photon resonances with transitions along the high symmetry axis between $\Gamma$ and $K$ where the highest valence and lowest conduction bands are flat and there is a high density of states; a broad structure between $2-3.5$~eV which originates from one-photon resonances with transitions at $K$ and along $\Gamma-K$ and two-photon resonances with transitions at higher energies. Note that we do not include spin-orbit coupling in Eq.~\eqref{tdbse_shf}. The latter is expected to split the lowest peak into two weaker sub-peaks~\cite{molina2013effect}, but to leave unaffected the second peak, the one observed experimentally~\cite{PhysRevB.87.201401}.  

Because of the inhomogeneity of the MoS$_2$ monolayer, the addition of crystal local field effects within the TDH strongly modifies the SHG  [Fig.~\ref{fg:SHMoS2} (b)]. 
In particular the main peak at $1.5$~eV merges with the plateau at $2$~eV while a peak appears around $3.3$~eV.
Finally, within the TDSHF [Fig.~\ref{fg:SHMoS2} (c)] the small shoulder around $1$~eV, below half of the $GW$ gap (1.25 eV, grey dotted vertical line in the figure), originates from two-photon resonances with the bound excitons around $2$~eV which are well visible in the experimental absorption spectra~\cite{PhysRevLett.105.136805}.  The main peak at about $1.5$~eV, present in the IPA spectrum but washed out by local field effects within the TDH, is restored by the electron-hole interaction and its intensity is two times larger than in the IPA case. This peak corresponds to a two-photon resonance with the bright exciton at 3~eV observed in the absorption spectrum~\cite{molina2013effect,PhysRevB.87.201401}. The calculated spectrum also shows a strong one-photon resonance with the same exciton at 3~eV.  

Interestingly the two-photon resonance with the bright exciton at 3 eV falls into the wavelength range of Ti:sapphire lasers. In fact it has been measured recently in different experiments~\cite{PhysRevB.87.201401,doi:10.1021/nl401561r,kumar2013second} reporting estimates for the SHG at 810 nm (1.53 eV) ranging over 4 orders of magnitude. In Fig.~\ref{fg:SHMoS2}(c) we compare the TDSHF calculated spectrum with the experimental measurements of Malard et al.~\cite{PhysRevB.87.201401} between $1.2-1.7$~eV, finding a good agreement for the position and shape of the peak at about $1.5$~eV. The calculated intensity though is one order of magnitude larger (about a factor 21) than the experimental estimate from both Refs.~\cite{PhysRevB.87.201401,doi:10.1021/nl401561r} (assuming an effective layer thickness of 6.2 \AA). As for h-BN, a quantitative comparison with experiment is however not possible: in the experiment the monolayer is deposited on quartz and the second harmonic signal measured with respect to the substrate while we calculate the SHG of the monolayer in vacuum. 
On the other hand our value is smaller by 2-3 orders of magnitude than the experimental estimate reported in Ref.~\cite{kumar2013second}. The same differences between theoretical and experimental SHG has been also reported recently by Trolle and coworkers~\cite{2013arXiv1310.0674T} that calculated the SHG of the MoS$_2$ monolayer in the IPA from a tight-binding band structure.
We note finally that our calculations predict the difference of 1 order of magnitude between the SHG in MoS$_2$ and h-BN at 810 nm (1.53 eV) as reported in Ref.~\cite{doi:10.1021/nl401561r}.

To summarize, we have shown that electron-hole interaction greatly enhances the SHG signal in 2D crystals with respect to the independent-particle picture. Specifically, for the h-BN monolayer one- and two-photons resonances with bound excitons produce strong signatures in the SHG spectrum with intensities two times larger than expected from the IPA. In MoS$_2$, though the shape of the spectrum is not strikingly modified by excitonic effects as for h-BN, the electron-hole interaction enhances, again by about 200\%, the SHG signal in the visible range with respect to the IPA. 
 
This finding may provide a spin-off for the quest of materials with high SHG. In fact---given that the SHG signal depends largely on the electron-hole interaction that in turn depends on the electronic screening---the SHG intensity can be tuned by changing the electronic screening. Then, it may be possible, as proposed in Ref.~\cite{gao2012artificially}, to engineer meta-materials with a high SHG by combining layers of different 2D crystals~\cite{gao2012artificially} so to change the electronic screening, and further enhance the electron-hole interaction effects.

As side finding, our results emphasize that it is critical for theoretical and computational approaches to accurately include electron-hole interaction, together with quasiparticle and local field effects, in order to predict non-linear optical response in low dimensional materials. In this regard, our recently proposed approach~\cite{nloptics2013,attaccalite} is quite promising as it imports into the very flexible real-time framework---apt to treat nonlinear optics---the combination of BSE+$GW$ successfully applied to the linear optical response of low-dimensional materials. 

\emph{Acknowledgments} The authors thanks L. Stella and J. Kohanoff for critical reading of the manuscript, and X. Blase for computational resources. M.G. acknowledges the Portuguese Foundation for Science and Technology for funding (PTDC/FIS/103587/2008) and support through the Ci\^encia 2008 program. Computing time has been provided by the national GENGI-IDRIS supercomputing centers, contract $n^o$ i2012096655.

\begin{thebibliography}{10}%
\makeatletter
\providecommand \@ifxundefined [1]{%
 \ifx #1\undefined \expandafter \@firstoftwo
 \else \expandafter \@secondoftwo
\fi
}%
\providecommand \@ifnum [1]{%
 \ifnum #1\expandafter \@firstoftwo
 \else \expandafter \@secondoftwo
\fi
}%
\providecommand \enquote [1]{``#1''}%
\providecommand \bibnamefont  [1]{#1}%
\providecommand \bibfnamefont [1]{#1}%
\providecommand \citenamefont [1]{#1}%
\providecommand\href[0]{\@sanitize\@href}%
\providecommand\@href[1]{\endgroup\@@startlink{#1}\endgroup\@@href}%
\providecommand\@@href[1]{#1\@@endlink}%
\providecommand \@sanitize [0]{\begingroup\catcode`\&12\catcode`\#12\relax}%
\@ifxundefined \pdfoutput {\@firstoftwo}{%
 \@ifnum{\z@=\pdfoutput}{\@firstoftwo}{\@secondoftwo}%
}{%
 \providecommand\@@startlink[1]{\leavevmode}%
 \providecommand\@@endlink[0]{}%
}{%
 \providecommand\@@startlink[1]{%
  \leavevmode
  \pdfstartlink
   attr{/Border[0 0 1 ]/H/I/C[0 1 1]}%
   user{/Subtype/Link/A<</Type/Action/S/URI/URI(#1)>>}%
  \relax
 }%
 \providecommand\@@endlink[0]{\pdfendlink}%
}%
\providecommand \url  [0]{\begingroup\@sanitize \@url }%
\providecommand \@url [1]{\endgroup\@href {#1}{\urlprefix}}%
\providecommand \urlprefix [0]{URL }%
\providecommand \Eprint[0]{\href }%
\@ifxundefined \urlstyle {%
  \providecommand \doi [1]{doi:\discretionary{}{}{}#1}%
}{%
  \providecommand \doi [0]{doi:\discretionary{}{}{}\begingroup
  \urlstyle{rm}\Url }%
}%
\providecommand \doibase [0]{http://dx.doi.org/}%
\providecommand \Doi[1]{\href{\doibase#1}}%
\providecommand \bibAnnote [3]{%
  \BibitemShut{#1}%
  \begin{quotation}\noindent
    \textsc{Key:}\ #2\\\textsc{Annotation:}\ #3%
  \end{quotation}%
}%
\providecommand \bibAnnoteFile [2]{%
  \IfFileExists{#2}{\bibAnnote {#1} {#2} {\input{#2}}}{}%
}%
\providecommand \typeout [0]{\immediate \write \m@ne }%
\providecommand \selectlanguage [0]{\@gobble}%
\providecommand \bibinfo [0]{\@secondoftwo}%
\providecommand \bibfield [0]{\@secondoftwo}%
\providecommand \translation [1]{[#1]}%
\providecommand \BibitemOpen[0]{}%
\providecommand \bibitemStop [0]{}%
\providecommand \bibitemNoStop [0]{.\EOS\space}%
\providecommand \EOS [0]{\spacefactor3000\relax}%
\providecommand \BibitemShut [1]{\csname bibitem#1\endcsname}%
\bibitem{doi:10.1021/nl903868w}%
  \BibitemOpen
  \bibfield{author}{%
  \bibinfo {author} {\bibfnamefont{A.}~\bibnamefont{Splendiani}}
  \emph{et~al.},\ }%
  \bibfield{journal}{%
  \bibinfo {journal} {Nano Letters}\ }%
  \textbf{\bibinfo {volume} {10}},\ \bibinfo {pages} {1271} (\bibinfo {year}
  {2010})%
  \bibAnnoteFile{NoStop}{doi:10.1021/nl903868w}%
\bibitem{PhysRevLett.105.136805}%
  \BibitemOpen
  \bibfield{author}{%
  \bibinfo {author} {\bibfnamefont{K.~F.}\ \bibnamefont{Mak}}, \bibinfo
  {author} {\bibfnamefont{C.}~\bibnamefont{Lee}}, \bibinfo {author}
  {\bibfnamefont{J.}~\bibnamefont{Hone}}, \bibinfo {author}
  {\bibfnamefont{J.}~\bibnamefont{Shan}},\ and\ \bibinfo {author}
  {\bibfnamefont{T.~F.}\ \bibnamefont{Heinz}},\ }%
  \bibfield{journal}{%
  \bibinfo {journal} {Phys. Rev. Lett.}\ }%
  \textbf{\bibinfo {volume} {105}},\ \bibinfo {pages} {136805} (\bibinfo {year}
  {2010})%
  \bibAnnoteFile{NoStop}{PhysRevLett.105.136805}%
\bibitem{watanabe}%
  \BibitemOpen
  \bibfield{author}{%
  \bibinfo {author} {\bibfnamefont{Y.}~\bibnamefont{Kubota}}, \bibinfo {author}
  {\bibfnamefont{K.}~\bibnamefont{Watanabe}}, \bibinfo {author}
  {\bibfnamefont{O.}~\bibnamefont{Tsuda}},\ and\ \bibinfo {author}
  {\bibfnamefont{T.}~\bibnamefont{Taniguchi}},\ }%
  \bibfield{journal}{%
  \bibinfo {journal} {Science}\ }%
  \textbf{\bibinfo {volume} {317}},\ \bibinfo {pages} {932} (\bibinfo {year}
  {2007})%
  \bibAnnoteFile{NoStop}{watanabe}%
\bibitem{Watanabe2004}%
  \BibitemOpen
  \bibfield{author}{%
  \bibinfo {author} {\bibfnamefont{K.}~\bibnamefont{Watanabe}}, \bibinfo
  {author} {\bibfnamefont{T.}~\bibnamefont{Taniguchi}},\ and\ \bibinfo {author}
  {\bibfnamefont{H.}~\bibnamefont{Kanda}},\ }%
  \bibfield{journal}{%
  \bibinfo {journal} {Nat Mater}\ }%
  \textbf{\bibinfo {volume} {3}},\ \bibinfo {pages} {404} (\bibinfo {year}
  {2004})%
  \bibAnnoteFile{NoStop}{Watanabe2004}%
\bibitem{doi:10.1021/nn403159y}%
  \BibitemOpen
  \bibfield{author}{%
  \bibinfo {author} {\bibfnamefont{G.}~\bibnamefont{Eda}}\ and\ \bibinfo
  {author} {\bibfnamefont{S.~A.}\ \bibnamefont{Maier}},\ }%
  \bibfield{journal}{%
  \bibinfo {journal} {ACS Nano}\ }%
  \textbf{\bibinfo {volume} {7}},\ \bibinfo {pages} {5660} (\bibinfo {year}
  {2013})%
  \bibAnnoteFile{NoStop}{doi:10.1021/nn403159y}%
\bibitem{molina2013effect}%
  \BibitemOpen
  \bibfield{author}{%
  \bibinfo {author} {\bibfnamefont{A.}~\bibnamefont{Molina-S\'anchez}},
  \bibinfo {author} {\bibfnamefont{D.}~\bibnamefont{Sangalli}}, \bibinfo
  {author} {\bibfnamefont{K.}~\bibnamefont{Hummer}}, \bibinfo {author}
  {\bibfnamefont{A.}~\bibnamefont{Marini}},\ and\ \bibinfo {author}
  {\bibfnamefont{L.}~\bibnamefont{Wirtz}},\ }%
  \bibfield{journal}{%
  \bibinfo {journal} {Phys. Rev. B}\ }%
  \textbf{\bibinfo {volume} {88}},\ \bibinfo {pages} {045412} (\bibinfo {year}
  {2013})%
  \bibAnnoteFile{NoStop}{molina2013effect}%
\bibitem{PhysRevLett.100.189701}%
  \BibitemOpen
  \bibfield{author}{%
  \bibinfo {author} {\bibfnamefont{L.}~\bibnamefont{Wirtz}}, \bibinfo {author}
  {\bibfnamefont{A.}~\bibnamefont{Marini}}, \bibinfo {author}
  {\bibfnamefont{M.}~\bibnamefont{Gr\"uning}}, \bibinfo {author}
  {\bibfnamefont{C.}~\bibnamefont{Attaccalite}}, \bibinfo {author}
  {\bibfnamefont{G.}~\bibnamefont{Kresse}},\ and\ \bibinfo {author}
  {\bibfnamefont{A.}~\bibnamefont{Rubio}},\ }%
  \bibfield{journal}{%
  \bibinfo {journal} {Phys. Rev. Lett.}\ }%
  \textbf{\bibinfo {volume} {100}},\ \bibinfo {pages} {189701} (\bibinfo {year}
  {2008})%
  \bibAnnoteFile{NoStop}{PhysRevLett.100.189701}%
\bibitem{doi:10.1021/nl401544y}%
  \BibitemOpen
  \bibfield{author}{%
  \bibinfo {author} {\bibfnamefont{M.}~\bibnamefont{Bernardi}}, \bibinfo
  {author} {\bibfnamefont{M.}~\bibnamefont{Palummo}},\ and\ \bibinfo {author}
  {\bibfnamefont{J.~C.}\ \bibnamefont{Grossman}},\ }%
  \bibfield{journal}{%
  \bibinfo {journal} {Nano Letters}\ }%
  \textbf{\bibinfo {volume} {13}},\ \bibinfo {pages} {3664} (\bibinfo {year}
  {2013})%
  \bibAnnoteFile{NoStop}{doi:10.1021/nl401544y}%
\bibitem{scholes2006excitons}%
  \BibitemOpen
  \bibfield{author}{%
  \bibinfo {author} {\bibfnamefont{G.~D.}\ \bibnamefont{Scholes}}\ and\
  \bibinfo {author} {\bibfnamefont{G.}~\bibnamefont{Rumbles}},\ }%
  \bibfield{journal}{%
  \bibinfo {journal} {Nature materials}\ }%
  \textbf{\bibinfo {volume} {5}},\ \bibinfo {pages} {683} (\bibinfo {year}
  {2006})%
  \bibAnnoteFile{NoStop}{scholes2006excitons}%
\bibitem{kumar2013second}%
  \BibitemOpen
  \bibfield{author}{%
  \bibinfo {author} {\bibfnamefont{N.}~\bibnamefont{Kumar}}, \bibinfo {author}
  {\bibfnamefont{S.}~\bibnamefont{Najmaei}}, \bibinfo {author}
  {\bibfnamefont{Q.}~\bibnamefont{Cui}}, \bibinfo {author}
  {\bibfnamefont{F.}~\bibnamefont{Ceballos}}, \bibinfo {author}
  {\bibfnamefont{P.~M.}\ \bibnamefont{Ajayan}}, \bibinfo {author}
  {\bibfnamefont{J.}~\bibnamefont{Lou}},\ and\ \bibinfo {author}
  {\bibfnamefont{H.}~\bibnamefont{Zhao}},\ }%
  \bibfield{journal}{%
  \bibinfo {journal} {Phys. Rev. B}\ }%
  \textbf{\bibinfo {volume} {87}},\ \bibinfo {pages} {161403} (\bibinfo {year}
  {2013})%
  \bibAnnoteFile{NoStop}{kumar2013second}%
\bibitem{doi:10.1021/nl401561r}%
  \BibitemOpen
  \bibfield{author}{%
  \bibinfo {author} {\bibfnamefont{Y.}~\bibnamefont{Li}} \emph{et~al.},\ }%
  \bibfield{journal}{%
  \bibinfo {journal} {Nano Letters}\ }%
  \textbf{\bibinfo {volume} {13}},\ \bibinfo {pages} {3329} (\bibinfo {year}
  {2013})%
  \bibAnnoteFile{NoStop}{doi:10.1021/nl401561r}%
\bibitem{PhysRevB.87.201401}%
  \BibitemOpen
  \bibfield{author}{%
  \bibinfo {author} {\bibfnamefont{L.~M.}\ \bibnamefont{Malard}}, \bibinfo {author} {\bibfnamefont{T.~V.}\ \bibnamefont{Alencar}},
  \bibinfo {author} {\bibfnamefont{A.~M.}\ \bibnamefont{Barboza}}, \bibinfo {author} {\bibfnamefont{K.~F.}\ \bibnamefont{Mak}},,\ and\ \bibinfo {author}
  {\bibfnamefont{A.~M.}~\bibnamefont{de Paula}},\ }%
  \bibfield{journal}{%
  \bibinfo {journal} {Phys. Rev. B}\ }%
  \textbf{\bibinfo {volume} {87}},\ \bibinfo {pages} {201401} (\bibinfo {year}
  {2013})%
  \bibAnnoteFile{NoStop}{PhysRevB.87.201401}%
\bibitem{Leitsmann2005}%
  \BibitemOpen
  \bibfield{author}{%
  \bibinfo {author} {\bibfnamefont{R.}~\bibnamefont{Leitsmann}}, \bibinfo
  {author} {\bibfnamefont{W.~G.}\ \bibnamefont{Schmidt}}, \bibinfo {author}
  {\bibfnamefont{P.~H.}\ \bibnamefont{Hahn}},\ and\ \bibinfo {author}
  {\bibfnamefont{F.}~\bibnamefont{Bechstedt}},\ }%
  \bibfield{journal}{%
  \bibinfo {journal} {Phys. Rev. B}\ }%
  \textbf{\bibinfo {volume} {71}},\ \bibinfo {pages} {195209} (\bibinfo {year}
  {2005})%
  \bibAnnoteFile{NoStop}{Leitsmann2005}%
\bibitem{Chang2002}%
  \BibitemOpen
  \bibfield{author}{%
  \bibinfo {author} {\bibfnamefont{E.~K.}\ \bibnamefont{Chang}}, \bibinfo
  {author} {\bibfnamefont{E.~L.}\ \bibnamefont{Shirley}},\ and\ \bibinfo
  {author} {\bibfnamefont{Z.~H.}\ \bibnamefont{Levine}},\ }%
  \bibfield{journal}{%
  \bibinfo {journal} {Phys. Rev. B}\ }%
  \textbf{\bibinfo {volume} {65}},\ \bibinfo {pages} {035205} (\bibinfo {year}
  {2001})%
  \bibAnnoteFile{NoStop}{Chang2002}%
\bibitem{PhysRevB.82.235201}%
  \BibitemOpen
  \bibfield{author}{%
  \bibinfo {author} {\bibfnamefont{E.}~\bibnamefont{Luppi}}, \bibinfo {author}
  {\bibfnamefont{H.}~\bibnamefont{H\"ubener}},\ and\ \bibinfo {author}
  {\bibfnamefont{V.}~\bibnamefont{V\'eniard}},\ }%
  \bibfield{journal}{%
  \bibinfo {journal} {Phys. Rev. B}\ }%
  \textbf{\bibinfo {volume} {82}},\ \bibinfo {pages} {235201} (\bibinfo {year}
  {2010})%
  \bibAnnoteFile{NoStop}{PhysRevB.82.235201}%
\bibitem{guo2005second}%
  \BibitemOpen
  \bibfield{author}{%
  \bibinfo {author} {\bibfnamefont{G.Y.}~\bibnamefont{Guo}}\ and\ \bibinfo
  {author} {\bibfnamefont{J.C.}~\bibnamefont{Lin}},\ }%
  \bibfield{journal}{%
  \bibinfo {journal} {Phys. Rev. B}\ }%
  \textbf{\bibinfo {volume} {72}},\ \bibinfo {pages} {075416} (\bibinfo {year}
  {2005})%
  \bibAnnoteFile{NoStop}{guo2005second}%
\bibitem{margulis2013optical}%
  \BibitemOpen
  \bibfield{author}{%
  \bibinfo {author} {\bibfnamefont{V.~A.}\ \bibnamefont{Margulis}}, \bibinfo
  {author} {\bibfnamefont{E.}~\bibnamefont{Muryumin}},\ and\ \bibinfo {author}
  {\bibfnamefont{E.}~\bibnamefont{Gaiduk}},\ }%
  \bibfield{journal}{%
  \bibinfo {journal} {Journal of Physics: Condensed Matter}\ }%
  \textbf{\bibinfo {volume} {25}},\ \bibinfo {pages} {195302} (\bibinfo {year}
  {2013})%
  \bibAnnoteFile{NoStop}{margulis2013optical}%
\bibitem{2013arXiv1310.0674T}%
  \BibitemOpen
  \bibfield{author}{%
  \bibinfo {author} {\bibfnamefont{M.~L.}\ \bibnamefont{{Trolle}}}, \bibinfo
  {author} {\bibfnamefont{G.}~\bibnamefont{{Seifert}}},\ and\ \bibinfo {author}
  {\bibfnamefont{T.~G.}\ \bibnamefont{{Pedersen}}},\ }%
  \enquote{\bibinfo {title} {{Theory of second harmonic generation in
  few-layered MoS2}},}\  (\bibinfo {year} {2013}),\ \bibinfo {note} {arXiv:
  1310.0674}%
  \bibAnnoteFile{NoStop}{2013arXiv1310.0674T}%
\bibitem{PhysRevB.80.165318}%
  \BibitemOpen
  \bibfield{author}{%
  \bibinfo {author} {\bibfnamefont{K.~S.}\ \bibnamefont{Virk}}\ and\ \bibinfo
  {author} {\bibfnamefont{J.~E.}\ \bibnamefont{Sipe}},\ }%
  \bibfield{journal}{%
  \bibinfo {journal} {Phys. Rev. B}\ }%
  \textbf{\bibinfo {volume} {80}},\ \bibinfo {pages} {165318} (\bibinfo {year}
  {2009})%
  \bibAnnoteFile{NoStop}{PhysRevB.80.165318}%
\bibitem{kirtman:1294}%
  \BibitemOpen
  \bibfield{author}{%
  \bibinfo {author} {\bibfnamefont{B.}~\bibnamefont{Kirtman}}, \bibinfo
  {author} {\bibfnamefont{F.~L.}\ \bibnamefont{Gu}},\ and\ \bibinfo {author}
  {\bibfnamefont{D.~M.}\ \bibnamefont{Bishop}},\ }%
  \bibfield{journal}{%
  \bibinfo {journal} {The Journal of Chemical Physics}\ }%
  \textbf{\bibinfo {volume} {113}},\ \bibinfo {pages} {1294} (\bibinfo {year}
  {2000})%
  \bibAnnoteFile{NoStop}{kirtman:1294}%
\bibitem{nloptics2013}%
  \BibitemOpen
  \bibfield{author}{%
  \bibinfo {author} {\bibfnamefont{C.}~\bibnamefont{Attaccalite}}\ and\
  \bibinfo {author} {\bibfnamefont{M.}~\bibnamefont{Gr\"uning}},\ }%
  \enquote{\bibinfo {title} {Nonlinear optics from ab-initio by means of the
  dynamical berry-phase},}\ \bibinfo {note} {ArXiv: 1309.4012}%
  \bibAnnoteFile{NoStop}{nloptics2013}%
\bibitem{takimoto:154114}%
  \BibitemOpen
  \bibfield{author}{%
  \bibinfo {author} {\bibfnamefont{Y.}~\bibnamefont{Takimoto}}, \bibinfo
  {author} {\bibfnamefont{F.~D.}\ \bibnamefont{Vila}},\ and\ \bibinfo {author}
  {\bibfnamefont{J.~J.}\ \bibnamefont{Rehr}},\ }%
  \bibfield{journal}{%
  \bibinfo {journal} {The Journal of Chemical Physics}\ }%
  \textbf{\bibinfo {volume} {127}},\ \bibinfo {pages} {154114} (\bibinfo {year}
  {2007})%
  \bibAnnoteFile{NoStop}{takimoto:154114}%
\bibitem{strinati}%
  \BibitemOpen
  \bibfield{author}{%
  \bibinfo {author} {\bibfnamefont{G.}~\bibnamefont{Strinati}},\ }%
  \bibfield{journal}{%
  \bibinfo {journal} {Rivista del nuovo cimento}\ }%
  \textbf{\bibinfo {volume} {11}},\ \bibinfo {pages} {1} (\bibinfo {year}
  {1988})%
  \bibAnnoteFile{NoStop}{strinati}%
\bibitem{RevModPhys.66.899}%
  \BibitemOpen
  \bibfield{author}{%
  \bibinfo {author} {\bibfnamefont{R.}~\bibnamefont{Resta}},\ }%
  \bibfield{journal}{%
  \bibinfo {journal} {Rev. Mod. Phys.}\ }%
  \textbf{\bibinfo {volume} {66}},\ \bibinfo {pages} {899} (\bibinfo {year}
  {1994})%
  \bibAnnoteFile{NoStop}{RevModPhys.66.899}%
\bibitem{souza_prb}%
  \BibitemOpen
  \bibfield{author}{%
  \bibinfo {author} {\bibfnamefont{I.}~\bibnamefont{Souza}}, \bibinfo {author}
  {\bibfnamefont{J.}~\bibnamefont{\'I\~niguez}},\ and\ \bibinfo {author}
  {\bibfnamefont{D.}~\bibnamefont{Vanderbilt}},\ }%
  \bibfield{journal}{%
  \bibinfo {journal} {Phys. Rev. B}\ }%
  \textbf{\bibinfo {volume} {69}},\ \bibinfo {pages} {085106} (\bibinfo {year}
  {2004})%
  \bibAnnoteFile{NoStop}{souza_prb}%
\bibitem{PhysRev.140.A1133}%
  \BibitemOpen
  \bibfield{author}{%
  \bibinfo {author} {\bibfnamefont{W.}~\bibnamefont{Kohn}}\ and\ \bibinfo
  {author} {\bibfnamefont{L.~J.}\ \bibnamefont{Sham}},\ }%
  \bibfield{journal}{%
  \bibinfo {journal} {Phys. Rev.}\ }%
  \textbf{\bibinfo {volume} {140}},\ \bibinfo {pages} {A1133} (\bibinfo {year}
  {1965})%
  \bibAnnoteFile{NoStop}{PhysRev.140.A1133}%
\bibitem{aulbur1999quasiparticle}%
  \BibitemOpen
  \bibfield{author}{%
  \bibinfo {author} {\bibfnamefont{W.~G.}\ \bibnamefont{Aulbur}}, \bibinfo
  {author} {\bibfnamefont{L.}~\bibnamefont{J{\"o}nsson}},\ and\ \bibinfo
  {author} {\bibfnamefont{J.~W.}\ \bibnamefont{Wilkins}},\ }%
  \bibfield{journal}{%
  \bibinfo {journal} {Solid State Physics}\ }%
  \textbf{\bibinfo {volume} {54}},\ \bibinfo {pages} {1} (\bibinfo {year}
  {1999})%
  \bibAnnoteFile{NoStop}{aulbur1999quasiparticle}%
\bibitem{attaccalite}%
  \BibitemOpen
  \bibfield{author}{%
  \bibinfo {author} {\bibfnamefont{C.}~\bibnamefont{Attaccalite}}, \bibinfo
  {author} {\bibfnamefont{M.}~\bibnamefont{Gr\"uning}},\ and\ \bibinfo {author}
  {\bibfnamefont{A.}~\bibnamefont{Marini}},\ }%
  \bibfield{journal}{%
  \bibinfo {journal} {Phys. Rev. B}\ }%
  \textbf{\bibinfo {volume} {84}},\ \bibinfo {pages} {245110} (\bibinfo {year}
  {2011})%
  \bibAnnoteFile{NoStop}{attaccalite}%
\bibitem{PhysRev.126.413}%
  \BibitemOpen
  \bibfield{author}{%
  \bibinfo {author} {\bibfnamefont{S.~L.}\ \bibnamefont{Adler}},\ }%
  \bibfield{journal}{%
  \bibinfo {journal} {Phys. Rev.}\ }%
  \textbf{\bibinfo {volume} {126}},\ \bibinfo {pages} {413} (\bibinfo {year}
  {1962})%
  \bibAnnoteFile{NoStop}{PhysRev.126.413}%
\bibitem{hbnparameters}%
  \BibitemOpen
  \bibinfo {note} {For the hexagonal-BN we used a lattice constant of
  $a=2.52~\AA$, and an inter-layer distance of 20 a.u. The electron-ion term
  was approximated using norm conserving pseudo-potentials and exchange
  correlation term using the local density approximation. The $GW$ correction
  of 3.3 eV was taken from
  Refs.~\cite{PhysRevLett.96.126104,PhysRevLett.100.189701}. The screened
  Coulomb interaction was calculated using 30 bands and a cutoff of 2 Ha on the
  dielectric matrix dimensions. SHG was obtained using 8 bands: 4 valence and 4
  conduction. Within IP and TDH we used a $40\times40\times1$ k-point grid,
  within TDSHF the optical spectra is dominated by a strongly bound exciton and
  a $14\times 14\times 1$ k-point grid is sufficient to converge the spectra.}%
  \bibAnnoteFile{Stop}{hbnparameters}%
\bibitem{mos2parameters}%
  \BibitemOpen
  \bibinfo {note} {For the MoS$_2$ layer we used the experimental lattice
  constant of the bulk MoS$_2$ as in Ref.~\cite{molina2013effect} and an
  inter-layer distance of 30 a.u. The electron-ion term was approximated using
  norm conserving pseudo-potentials and exchange correlation term using the
  Perdew, Burke and Ernzerhof functional [Phys. Rev. Lett. 77, 3865 (1996)].
  The $GW$ correction of $0.72~eV$ was taken from Ref.~\cite{molina2013effect}.
  The screened Coulomb interaction was calculated using 100 bands and a cutoff
  of 2 Ha on the dielectric matrix dimensions. SHG was obtained using a
  $21\times21\times1$ k-point grid, 18 bands within IPA and TDH, and bands
  between the 3$^\text{rd}$ and the 16$^\text{th}$ for the TDSHF.}%
  \bibAnnoteFile{Stop}{mos2parameters}%
\bibitem{intequation}%
  \BibitemOpen
  \bibinfo {note} {Calculations of $H_\kk^0$ are performed with {\sc
  Abinit}~\cite{abinit}, while Eq.~\eqref{tdbse_shf} is solved using a
  development version of the {\sc Yambo} code~\cite{yambo}, where also
  Eq.~\eqref{berryP} have been implemented. Equation~\eqref{tdbse_shf} is
  numerically integrated with a time-step of $\Delta t = 0.0025$~fs, that
  guarantees accuracy and stable results. In order to reproduce experimental
  conditions we use a laser intensity of $I = 500~$kW/cm$^2$. In Eq.~\eqref{tdbse_shf} we add a
  dephasing term with $\tau=6$~fs to simulate a finite
  broadening of about $0.2~eV$~\cite{nloptics2013}, and propagate for
  $55$~fs for each laser frequency.}%
  \bibAnnoteFile{Stop}{intequation}%
\bibitem{PhysRevLett.96.126104}%
  \BibitemOpen
  \bibfield{author}{%
  \bibinfo {author} {\bibfnamefont{L.}~\bibnamefont{Wirtz}}, \bibinfo {author}
  {\bibfnamefont{A.}~\bibnamefont{Marini}},\ and\ \bibinfo {author}
  {\bibfnamefont{A.}~\bibnamefont{Rubio}},\ }%
  \bibfield{journal}{%
  \bibinfo {journal} {Phys. Rev. Lett.}\ }%
  \textbf{\bibinfo {volume} {96}},\ \bibinfo {pages} {126104} (\bibinfo {year}
  {2006})%
  \bibAnnoteFile{NoStop}{PhysRevLett.96.126104}%
\bibitem{hughes1996calculation}%
  \BibitemOpen
  \bibfield{author}{%
  \bibinfo {author} {\bibfnamefont{J.~L.~P.}\ \bibnamefont{Hughes}}\ and\ \bibinfo
  {author} {\bibfnamefont{J.~E.}~\bibnamefont{Sipe}},\ }%
  \bibfield{journal}{%
  \bibinfo {journal} {Phys. Rev. B}\ }%
  \textbf{\bibinfo {volume} {53}},\ \bibinfo {pages} {10751} (\bibinfo {year}
  {1996})%
  \bibAnnoteFile{NoStop}{hughes1996calculation}%
\bibitem{gao2012artificially}%
  \BibitemOpen
  \bibfield{author}{%
  \bibinfo {author} {\bibfnamefont{G.}~\bibnamefont{Gao}} \emph{et~al.},\ }%
  \bibfield{journal}{%
  \bibinfo {journal} {Nano letters}\ }%
  \textbf{\bibinfo {volume} {12}},\ \bibinfo {pages} {3518} (\bibinfo {year}
  {2012})%
  \bibAnnoteFile{NoStop}{gao2012artificially}%
\bibitem{abinit}%
  \BibitemOpen
  \bibfield{author}{%
  \bibinfo {author} {\bibfnamefont{X.}~\bibnamefont{Gonze}} \emph{et~al.},\ }%
  \bibfield{journal}{%
  \bibinfo {journal} {Comp. Mat. Sci.}\ }%
  \textbf{\bibinfo {volume} {25}},\ \bibinfo {pages} {478 } (\bibinfo {year}
  {2002})%
  \bibAnnoteFile{NoStop}{abinit}%
\bibitem{yambo}%
  \BibitemOpen
  \bibfield{author}{%
  \bibinfo {author} {\bibfnamefont{A.}~\bibnamefont{Marini}}, \bibinfo {author}
  {\bibfnamefont{C.}~\bibnamefont{Hogan}}, \bibinfo {author}
  {\bibfnamefont{M.}~\bibnamefont{Gruning}},\ and\ \bibinfo {author}
  {\bibfnamefont{D.}~\bibnamefont{Varsano}},\ }%
  \bibfield{journal}{%
  \bibinfo {journal} {Comp. Phys. Comm.}\ }%
  \textbf{\bibinfo {volume} {180}},\ \bibinfo {pages} {1392} (\bibinfo {year}
  {2009})%
  \bibAnnoteFile{NoStop}{yambo}%
\end{thebibliography}
%

\end{document}